\documentclass{article}
\usepackage{LaThuileFPSpro}
\usepackage{graphicx}
\begin{document}

\title{ 
  PRECISION DETERMINATION OF THE TOP QUARK MASS
  }
\author{
  Pedro A. Movilla Fern\'andez        \\
  {\em Lawrence Berkeley National Laboratory,} \\ 
  {\em 1 Cyclotron Rd., Berkeley, CA 94720, U.S.A.} \\[2mm]
  For the CDF and D\O\ Collaborations
  }
\maketitle

\baselineskip=11.6pt

\newcommand{\via} 	{{\it via}}
\newcommand{\eg}  	{{\it e.g.}}
\newcommand{\Eg}  	{{\it E.g.}}
\newcommand{\vs}   	{{\it vs.}}
\newcommand{\etal}   	{{\it et al.}}
\newcommand{\etc} 	{{\it etc.}}
\newcommand{\ie}  	{{\it i.e.}}
\newcommand{\Ie}  	{{\it I.e.}}

\newcommand{\MET} 	{\ensuremath{{\not}{E_\mathrm{T}}}}
\newcommand{\rs}   	{\ensuremath{\sqrt{s}}}
\newcommand{\MH}    	{\ensuremath{M_\mathrm{H}}}
\newcommand{\mH}    	{\ensuremath{m_\mathrm{H}}}
\newcommand{\MW}    	{\ensuremath{M_{W}}}
\newcommand{\mW}    	{\ensuremath{m_{W}}}
\newcommand{\Mt}    	{\ensuremath{M_{t}}}
\newcommand{\mt}    	{\ensuremath{m_{t}}}
\newcommand{\mtbar}	{\ensuremath{m_{\bar{t}}}}
\newcommand{\Mtop}	{\ensuremath{M_\mathrm{top}}}
\newcommand{\mtop}	{\ensuremath{m_\mathrm{top}}}
\newcommand{\DJES}	{\ensuremath{\Delta_\mathrm{JES}}}
\newcommand{\fJES}	{\ensuremath{f_\mathrm{JES}}}
\newcommand{\JES}	{\ensuremath{\mathrm{JES}}}
\newcommand{\fpdf}	{\ensuremath{f_\mathrm{PDF}}}
\newcommand{\dd} 	{\ensuremath{\mathrm{d}}}
\newcommand{\tbbar} 	{\ensuremath{{t\bar{b}}}}
\newcommand{\ttbar} 	{\ensuremath{{t\bar{t}}}}
\newcommand{\ppbar}	{\ensuremath{p\overline{p}}}
\newcommand{\bbbar}     {\ensuremath{{b\bar{b}}}}
\newcommand{\ccbar}     {\ensuremath{{c\bar{c}}}}
\newcommand{\qqbar}     {\ensuremath{{q\bar{q}}}}
\newcommand{\qqbarprime}{\ensuremath{{q\bar{q}^\prime}}}
\newcommand{\wplus}     {\ensuremath{{W^+}}}
\newcommand{\wminus}    {\ensuremath{{W^-}}}
\newcommand{\WbWb}  	{\ensuremath{W^+b}\ensuremath{W^-\bar{b}}}
\newcommand{\pt}        {\ensuremath{p_\mathrm{t}}}
\newcommand{\Et}        {\ensuremath{E_\mathrm{t}}}
\newcommand{\pT}        {\ensuremath{p_\mathrm{T}}}
\newcommand{\ET}        {\ensuremath{E_\mathrm{T}}}
\newcommand{\pl}        {\ensuremath{p_\mathrm{l}}}
\newcommand{\bm}[1]     {\mbox{\boldmath\ensuremath{#1}}}
\newcommand{\ra}        {\ensuremath{\rightarrow}}
\newcommand{\calO}      {\ensuremath{\mathcal{O}}}
\newcommand{\chisq}     {\ensuremath{\chi^2}}

\newcommand{\gevcc}     {\ensuremath{\mathrm{GeV}/c^2}}
\newcommand{\gevc}     	{\ensuremath{\mathrm{GeV}/c}}
\newcommand{\gev}       {\ensuremath{\mathrm{GeV}}}
\newcommand{\mev}       {\ensuremath{\mathrm{MeV}}}
\newcommand{\invpb}     {\ensuremath{\mathrm{pb}^{-1}}}
\newcommand{\invfb}     {\ensuremath{\mathrm{fb}^{-1}}}
\newcommand{\invnb}     {\ensuremath{\mathrm{nb}^{-1}}}
\newcommand{\invcms}    {\ensuremath{\mathrm{cm}^{-2}\mathrm{s}^{-1}}}
\newcommand{\degs}	{\mbox{$^{\circ}$}}
\newcommand{\gsim}	{\mbox{\small$\stackrel{>}{\sim}$\normalsize}}
\newcommand{\lsim}	{\mbox{\small$\stackrel{<}{\sim}$\normalsize}}

\newcommand{\nat}[1]{Nature {\bf #1}}
\newcommand{\prl}[1]{Phys. Rev. Lett. {\bf #1}}
\newcommand{\prev}[1]{Phys. Rev. {\bf #1}}
\newcommand{\prd}[1]{Phys. Rev. D {\bf #1}}
\newcommand{\zs}[1]{Z. Phys. {\bf #1}}
\newcommand{\ncim}[1]{Nuovo Cim. {\bf #1}}
\newcommand{\plet}[1]{Phys. Lett. {\bf #1}}
\newcommand{\prep}[1]{Phys. Rep. {\bf #1}}
\newcommand{\rmp}[1]{Rev. Mod. Phys. {\bf #1}}
\newcommand{\nphy}[1]{Nucl. Phys. {\bf #1}}
\newcommand{\nim}[1]{Nucl. Instrumen. Meth. {\bf #1}}

\begin{abstract}
  The CDF and D\O\ collaborations have updated their measurements of
  the mass of the top quark using proton-antiproton collisions at
  \rs=1.96\,TeV produced at the Tevatron. The uncertainties in each of
  the of top-antitop decay channels have been reduced. The new
  Tevatron average for the mass of the top quark based on about
  1\,\invfb\ of data per experiment is 170.9$\pm$1.8\,\gevcc.
\end{abstract}

\vspace*{-85mm}
\begin{flushright}
 FERMILAB-CONF-07-138-E
\end{flushright}
\vspace*{+85mm}

\bigskip
\bigskip
\bigskip
\bigskip
\bigskip
\bigskip


\newpage

\section{Introduction} \label{sec:intro}

The discovery of the top quark by the CDF and D\O\ collaborations
1995\cite{bib:topdiscovery} has marked the beginning of a successful
physics program at the Tevatron. The mass of the top quark (\Mt) is a
fundamental parameter of the Standard Model (SM), but more
importantly, its surprisingly high value gives the top quark
particular relevance in the calculation of other SM
parameters. Electroweak corrections to the $W$ propagator introduce a
quadratic dependence of the $W$ boson mass (\MW) on \Mt. \MW\ is also
expected to depend logarithmically on the mass of the
long-hypothesized but still unobserved Higgs boson (\MH). Thus, a
precision measurement of \Mt\ and \MW\ provides a mean to impose a
constraint to \MH. The presence of further loop corrections in which
heavy unknown particles are involved might lead to signatures beyond
the SM. The Yukawa coupling to the Higgs field of $\mathcal{O}(1)$
indicates that the top quark might play a special role in the
mechanism of electroweak symmetry breaking.

The determination of the top quark mass is therefore a very active
topic in Tevatron Run-II. The top quark mass has been measured in all
$\ttbar$ decay topologies with increasing precision.  Improvements are
based on the performance of the Tevatron, the better understanding of
the detectors, and particularly on innovative analysis
techniques. Here we report on the state-of-the-art of CDF and D\O\
measurements based on up to 1\,\invfb\ of analyzed data per
experiment.

\section{Experimental Challenges} \label{sec:exp}

The upgraded Tevatron complex started in 2001 to produce collisions of
protons and antiprotons at \rs\,=\,1.96\,TeV with steadily increasing
instantaneous luminosities up to a record of
3$\times$10$^{32}\,\invcms$. The CDF and D\O\
experiments\cite{bib:det} have integrated a luminosity of about
2\,\invfb\ each, the projected goal for the end of Run-II is
4-8\,\invfb. Both experiments are multipurpose detectors which cover
the interaction points almost hermetically.  The inner volumes contain
precision tracking systems and silicon vertex detectors embedded in a
solenoidal magnetic field. The magnets are surrounded by
electromagnetic and hadronic calorimeters. The outermost parts consist
of muon systems for the detection of penetrating particles. The
experiments are running with a data taking efficiency of better than
80\%.

At Tevatron energies, SM top quarks are mainly produced in pairs
through quark-antiquark annihilation (85\%) and gluon-gluon fusion
(15\%). The theoretical \ttbar-prodution cross section is
7.8$\pm$1.0\,pb\cite{bib:top-xsec} for a top quark mass of
170\,\gevcc, which corresponds to approximately one top quark pair
produced in $10^{10}$ inelastic collisions. The top quark does not
hadronize and promptly undergoes the transition $t\to Wb$ with a
branching ratio of BR$\sim$100\%. The event signature is thus defined
by the decay modes of the two $W$ bosons. We distinguish the
``di-lepton'' channel, $\ttbar\to(l_1^+\nu_{1}
b)(l_2^-\bar{\nu_{2}}\bar{b})$ (5\% fraction), the ``lepton-jets''
channel (30\% fraction), $\ttbar\to(q_1\bar{q_2}
b)(l^-\bar{\nu}\bar{b})$, and the ''all-jets'' channel,
$\ttbar\to(q_1\bar{q_2} b)(q_3\bar{q_4}\bar{b})$ (44\% fraction),
where the $q_i$'s stand for quarks and $l$ denote an electron $e$ or a
muon $\mu$ ($\tau$'s are usually ignored).%
\footnote{ \scriptsize A review of 
top quark property measurements is given by M.~Weber in these
proceedings.}

Top quark analyses require full detector capabilities. The measurement
of the leptonic $W$ decay modes relies on the clean identification of
electrons and muons. The hermeticity of the calorimeter is essential
for the partial reconstruction of the momentum of undetectable
neutrinos through the measurement of the missing transverse energy.
The reconstruction of the primary quarks from \ttbar\ decays involves
the accurate measurement of calorimeter energy deposits and their
appropriate clustering into jets. Quark flavor information is provided
by vertex detectors via the reconstruction of displaced vertices
consistent with long-lived $b$-hadrons, which are present in all decay
modes. The ``$b$-tagging'' is crucial to reduce background
contributions and the number of possible jet-quark assignments.

Top quark measurements critically depend on the accurate knowledge of
the jet energy scale (JES), which incorporates corrections of the raw
jet energies for physics and instrumental effects as well as for jet
definition artefacts. The JES is currently known {\em a priori} to a
level of 2-3\% for jets typical in \ttbar\ events\cite{bib:JES} and
constitutes the dominant source of systematic uncertainties.

\section{Measurement Techniques} \label{sec:tech}

The mass extraction techniques employed by CDF and D\O\ can be
subdivided into two categories. The {\em Template Method} is based on
the evaluation of one observable per event correlated with the top
quark mass \Mt, and a comparison of simulated distributions of this
observable (''templates'') with varying \Mt\ with the data.
Typically, some kind of reconstructed top quark mass \mt\ is taken,
for example the output of the kinematic fit of a \ttbar\ hypothesis to
the event. Recent analyses have introduced the JES as a second
template variable using distributions of the invariant di-jet mass
\mW\ of the hadronically decaying $W$ boson. The \mt\ and \mW\
distributions provide two-dimensional sample likelihoods which allow a
simultaneous determination of \Mt\ and the JES {\em in situ}. The
Template Method is computationally simple, but it uses limited event
information by evaluating just one or two numbers per event, and it
treats well and badly reconstructed events equally. Refined Template
Method analyses therefore apply weights to the events using further
kinematic information.

The {\em Matrix Element Method} enhances the mass information by
exploring the SM predictions for top quark dynamics. For each event, a
probability density curve $\mathcal{P}(\Mt)$ is extracted, which
expresses the quality of the agreement of the event with a signal or
background process as a function of \Mt. The per-event probabilities
are multiplied, and the maximum position of the resulting curve gives
the most likely value for \Mt\ for the whole signal candidate sample.
Recent measurements have extended the technique to allow the JES be
re-adjusted {\em in situ} using the invariant mass of the $W$ boson as
a reference:
\vspace*{-1.5mm}
\begin{equation} \label{eq:event-prob}
  \mathcal{P}_\ttbar(x|\Mt,\JES) \propto
\sum_\mathrm{comb} \int 
   \dd \sigma_\ttbar(y|\Mt)
   \dd q_1\dd q_2 f(q_1)f(q_2)
   w(x|y,\JES)\;.
\vspace*{-1.5mm}
\end{equation}
$\dd \sigma_\ttbar$ denotes the differential
\ttbar\ cross-section (using a tree level matrix element)
for a configuration of parton level momenta $y$, given \Mt, and
contains all integration details for the six-body phase
space. $f(q_i)$ is the proton-parton density function for given
momenta $q_i$ of the two incoming quarks. The transfer functions
$w(x|y,\JES)$ are probabilities of a set of variables $x$ (\eg\
transverse jet momenta) to be measured given a set of parton level
quantities $y$ (\eg\ quark momenta) and a shift of the JES from its
{\em a priori} known value. A JES hypothesis yielding to a $W$ mass
which is inconsistent with the known $W$ mass and width penalizes the
event probability. The transfer functions account for hadronization
effects and detector resolution. The sum usually goes over all
possible jet-quark permutations and neutrino solutions. The background
probabilities are calculated analog to Eq.\,(\ref{eq:event-prob}) but
have no $\Mt$ dependence.

Since the method buys its increased statistical power by CPU-intensive
numerical integrations, simplifying assumptions must be made in the
interest of computational tractability. Lepton momenta and jet angles
are often treated as exactly measured quantities, and only the
probability density shapes of the dominant background types are
calculated. Due to the various approximations, the method must be
calibrated using the behavior of fully simulated Monte Carlo (MC)
samples with known value for \Mt.

Both methods depend on trustworthy physics event generators and
detector simulations.  The {\em in-situ} technique has the advantage
that the largest part of the JES uncertainty becomes a statistical
component of the top quark mass uncertainty, which thus will scale
down as more luminosity is collected.

\section{Measurements in the Lepton-Jets Channel} \label{sec:lepjets}

The lepton-jets channel is viewed as a good compromise between all
decay modes because it has a reasonable branching fraction and a good
S/B ratio between $\sim$0.2-10, dependent on the $b$-tag requirement.
The final state is characterized by well defined kinematics with
moderate combinatorial quark-jet ambiguity. There are twelve ways to
assign jets to quarks if no $b$-tag information is used, and six (two)
possibilities in case of one (two) $b$-tags (ignoring the physically
equivalent permutations of the quarks from the $W$ boson). The number
of kinematic solutions doubles due to the twofold ambiguity of the
neutrino longitudinal momentum.

The event selection of both experiments usually requires one well
contained electron or muon candidate with transverse momentum
$\pT>20\,\gevc$, a sizable amount of missing transverse energy
$\MET>20\,\gev$ to account for the neutrino, and at least four jets
with $\ET\ge$15(20)\,\gev\ at CDF (D\O). Matrix element measurements
are restricted to events with exactly four jets, in order to match the
predicted final state partons, whereas template based analyses also
allow events with sub-leading jets to pass the selection.  Various
analyses subdivide the data into disjoint samples with different
$b$-tag cuts in order to handle statistical power against sample
purity. The background of this channel mainly consists of $W$+jets
final states (\eg\ $Wb\bar{b}q\bar{q}$, $Wq\bar{q}q\bar{q}$ with fake
$b$-tags, \etc) and QCD multi-jets events in which jets are
misidentified as leptons.

\begin{figure}[!t]
 \begin{center} 
 \includegraphics[width=.48\textwidth, height=55mm]{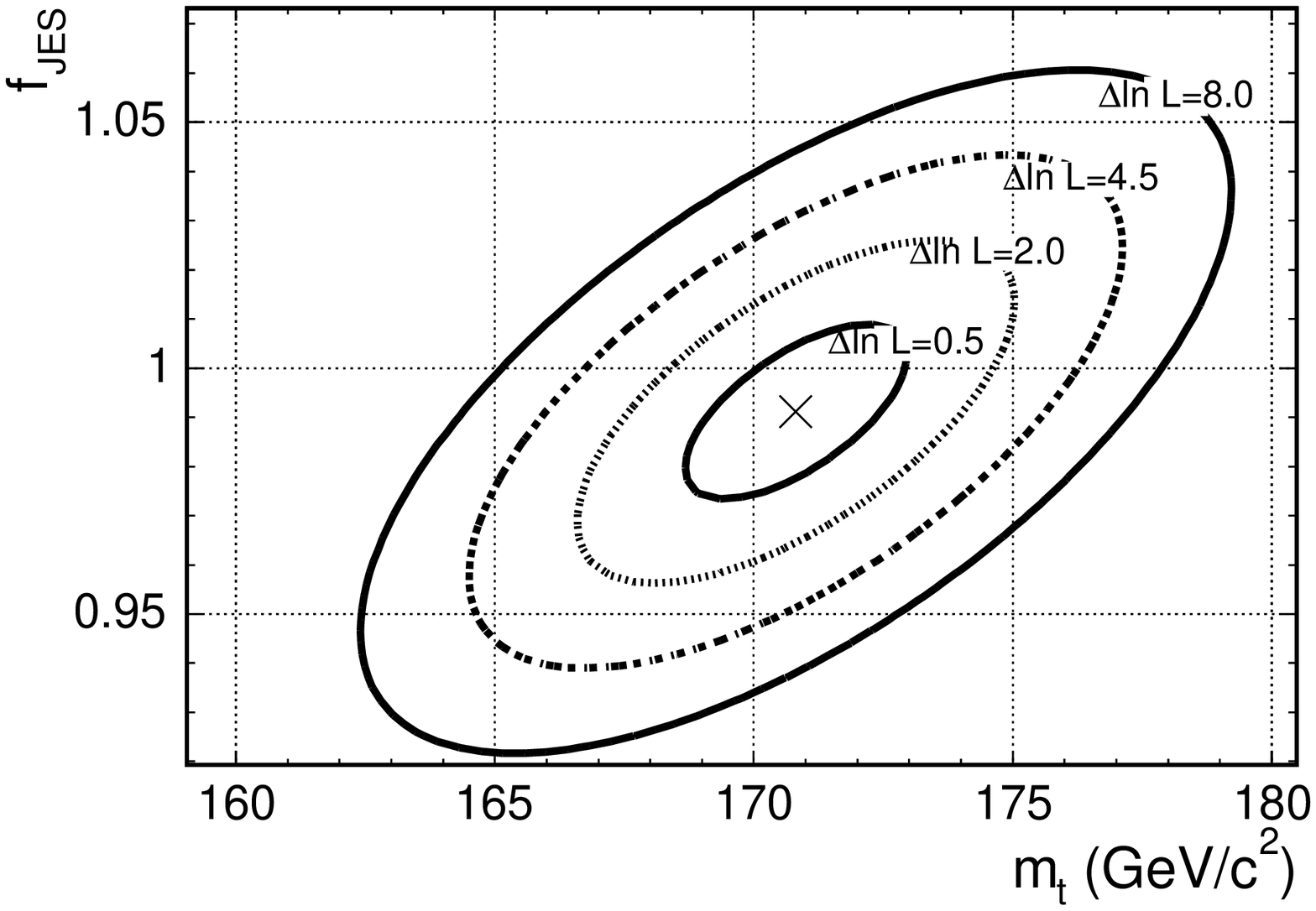}\hspace*{1mm}
 \includegraphics[width=.48\textwidth, height=55mm]{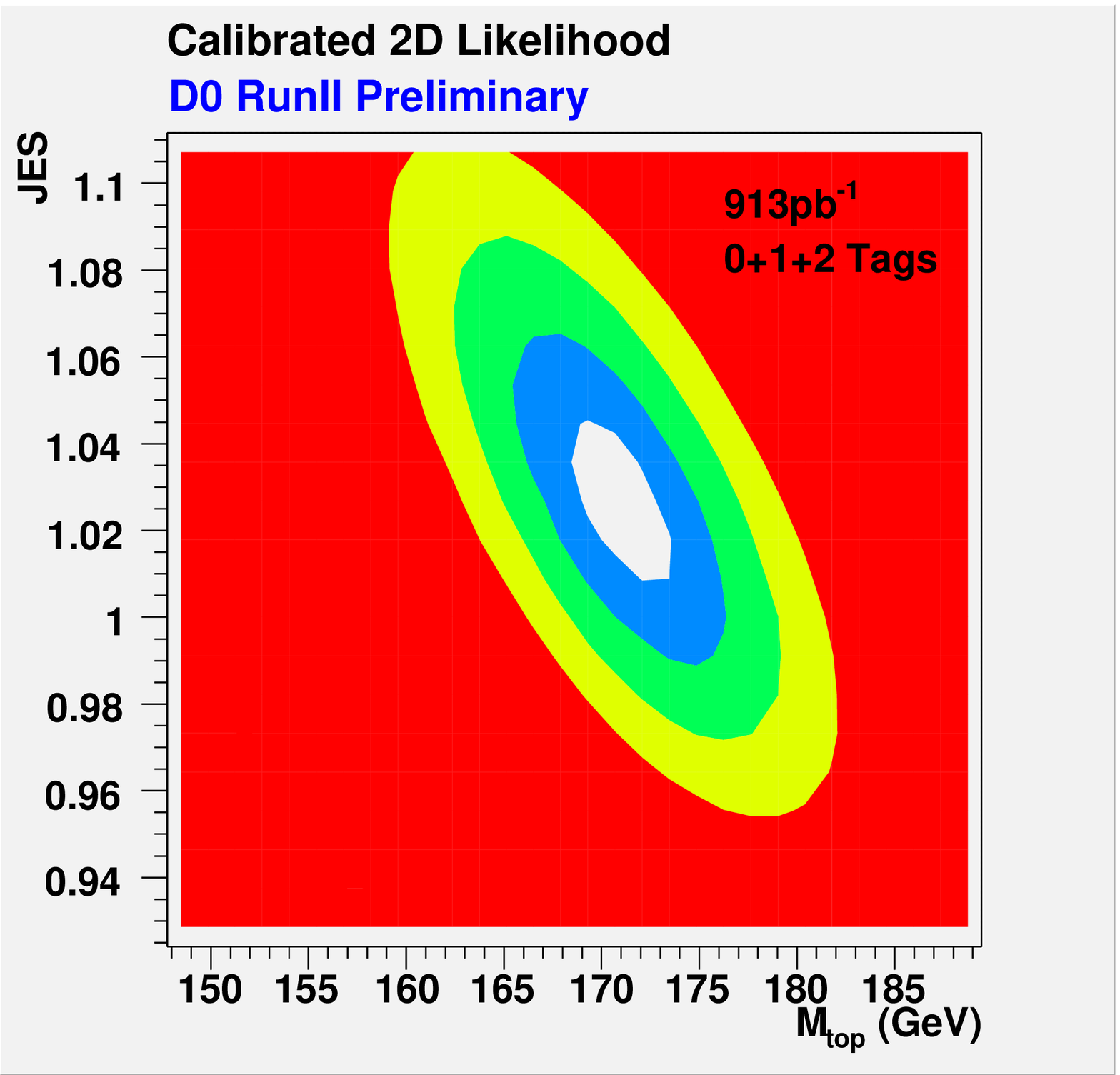} 
  \vspace*{-2mm}
  \caption{\it Likelihood contours obtained using the Matrix Element
  Method in the lepton-jets channel. Left: CDF result with
  955pb$^{\mathit{-1}}$; Right: D\O\ result with 913pb$^{\mathit{-1}}$
  (shown for $e$+jets events only).
   \vspace*{-6mm}
\label{fig:mtop_lepjets} }
 \end{center}
\end{figure}

Both experiments obtain the most precise results using the Matrix
Element Method with {\em in situ} JES calibration. CDF has
analyzed\cite{bib:CDF-LJ-ME} a data sample of 955\,\invpb\ and found
167 $b$-tagged candidate events ($22\pm8$ expected background).  The
signal ($\mathcal{P}_\ttbar$) and background probability densities
($\mathcal{P}_{W+\mathrm{jets}}$) are calculated similarly to
Eq.\,(\ref{eq:event-prob}) using a \ttbar\ leading order matrix
element and a MC based parametrization for the $W$+jets process (which
is also found to adequately describe the QCD multi-jets
probabilities).  A sample likelihood
\vspace*{-1mm}
\begin{equation} \label{eq:likelihood}
\mathcal{L}(\Mt,\JES)\propto \prod_i^\mathrm{\#events}
[f_t\mathcal{P}_\ttbar^{(i)}(\Mt,\JES)+(1-f_t)\mathcal{P}_{W+\mathrm{jets}}^{(i)}(\JES)]
\vspace*{-1mm}
\end{equation}
is used to extract simultaneously \Mt\ and the JES. The signal
fraction $f_t$ is allowed to float. Fig.\,\ref{fig:mtop_lepjets}
(left) shows the resulting likelihood contours. The analysis yields
$\Mt=170.8\pm2.2^\mathrm{(stat.+JES)}\pm1.4^\mathrm{(syst.)}\,\gevcc=170.8\pm2.6\,\gevcc$,
where the statistical component (stat.+JES) includes an uncertainty of
$1.5\,\gevcc$ due to the JES. With a relative uncertainty of 1.5\%,
this result constitutes the most precise measurement to date.

D\O\ has reported matrix element analyses\cite{bib:D0-LJ-ME} based on
913\,\invpb\ and 507 candidate events with $\ge 0$ $b$-tags
($373\pm39$ estimated background).  Two measurements are performed,
one which uses the $b$-tagger to assign weights to the jet-quark
assignments in the event probabilities, and a second one which focuses
on event topology only.  The $b$-tagging analysis calculates
individual likelihoods similarly to Eq.\,(\ref{eq:likelihood}) for
three subsamples with 0, 1 and $\ge2$ $b$-tags, and joins them using
individually optimized values for $f_t$.  Fig.\,\ref{fig:mtop_lepjets}
(right) shows the corresponding overall likelihood extracted from an
electron-jets subsample.  Differently from CDF, a JES prior is used,
and also the finite resolution of the electron and muon momentum is
considered in the transfer functions. The result obtained is
$\Mt=170.5\pm2.4^\mathrm{(stat.+JES)}\pm1.2^\mathrm{(syst.)}\,\gevcc=170.5\pm2.7\,\gevcc$
(1.6\% precision), where the uncertainty from the JES is
$1.6\,\gevcc$. The result of the pure topological analysis is
$\Mt=170.5\pm2.5\mathrm{(stat.+JES)}\pm1.4^\mathrm{(syst.)}\,\gevcc=170.5\pm2.9\,\gevcc$
(1.7\% precision).  The $b$-tagging analysis is the most precise D\O\
measurement and in excellent agreement with the CDF result.

The {\em in situ} JES calibration technique was pioneered by CDF and
originally used in Template Method analyses, of which the most recent
one\cite{bib:CDF-LJ-TM} uses a data set of 680\,\invpb. Four exclusive
samples with different S/B ratio and sensitivity to \Mt\ are selected
according to different $b$-tag requirements and jet \ET\ cuts. For
each sample, templates for \Mt\ and the JES are formed using the
reconstructed top quark mass (corresponding to the quark-jet
assignment with the lowest \chisq) and the invariant $W$ di-jet mass,
which are then compared to the data using an unbinned likelihood.  A
cut to the \chisq\ in addition to the standard selection ensures that
only well reconstructed events are considered.  Using 360 selected
candidate events ($97\pm23$ background from a constrained fit) the
result obtained is
$\Mt=173.4\pm2.5^\mathrm{(stat.+JES)}\pm1.3^\mathrm{(syst.)}\,\gevcc=173.4\pm2.8\,\gevcc$
(1.6\% precision), which is compatible with the matrix element
result. The JES contribution to the statistical uncertainty is
$1.8\,\gevcc$. The {\em in situ} calibration reduced the {\em a
priori} JES uncertainty by about 40\%.

CDF has recently demonstrated that the Template Method can be improved
by combining the kinematic top quark mass solutions of the three best
quark-jet combinations\cite{bib:CDF-LJ-ThreeBest}. The analysis
addresses the problem that the smallest \chisq\ corresponds to the
correct association in less than 50\% of the time. For each event, the
three solutions are combined considering their correlations.  A
analysis of 1030\,\invpb\ of data containing 645 candidates ($\ge0$
$b$-tags) yields
$\Mt=168.9\pm2.2^\mathrm{(stat.)}\pm4.2^\mathrm{(syst.)}\,\gevcc=168.9\pm4.7\,\gevcc$
(2.8\% precision), the best result achieved in this channel without
{\em in situ} JES calibration.

The D\O\ collaboration has employed the ``Ideogram Method'' (extended
by {\em in situ} JES calibration) for the first time in the
lepton-jets channel using a 425\,\invpb\ data
sample\cite{bib:DO-LJ-Ideogram} with 230 candidate events ($123\pm15$
expected background). Instead of evaluating matrix elements, the
analysis uses the outcome of a kinematic fitter, $b$-tagging
information and a multivariate S/B discriminant to extract per-event
probability densities similarly to Eq.\,(\ref{eq:likelihood}). The
signal probability $\mathcal{P}_\ttbar$ considers all possible
jet-quark permutations, which are weighted according to fit quality
and compatibility with $b$-tag information. The shape of the \Mt\
distribution of the correct permutation is given by a relativistic top
quark Breit Wigner function convoluted with an experimental Gaussian
resolution function, whereas the shape corresponding to the wrong
permutation as well for the background probability density
$\mathcal{P}_{W+\mathrm{jets}}$ is derived using appropriate MC
simulations. The result obtained is
$\Mt=173.7\pm4.4^\mathrm{(stat.+JES)}\ ^{+2.1}_{-2.0}\
^\mathrm{(syst.)}\,\gevcc$ (2.8\% precision).

CDF has performed further Template Method analyses in the lepton-jets
channel which currently lack statistical precision but are important
in the long run because they are aiming at establishing measurements
with different inherent systematics. Here we mention the ``Decay
Length Technique''\cite{bib:CDF-LJ-DecayLength}, which uses the
transverse distance of a jet's secondary vertex from the primary
vertex as a template variable. The method is motivated by the
expectation that $b$-hadrons from top quark decays are boosted and
thus correlated with \Mt. The analysis solely relies on tracking
information and has no JES dependence. Using 375 signal candidates
with 456 $b$-tagged jets found in a 695\,\invpb\ sample with at least
three jets per event yields $\Mt=180.7^{+15.5}_{-13.4}\
^\mathrm{(stat.)}\pm8.6^\mathrm{(syst.)}\,\gevcc$.  Despite the low
statistical precision, the method has proven its practicability and
can make significant contributions at LHC.

\section{Measurements in the Di-Lepton Channel} \label{sec:dilep}

The di-lepton channel provides pure signal samples but suffer from a
poor branching ratio of about 5\%. Experimentally, the event
kinematics is under-constrained due to the presence of two neutrinos
and the availability of just one missing \ET\ observable.
Template-based analyses therefore assume values for certain variables
(\eg\ the neutrino $\eta$) in order to extract a solution for the top
quark mass, and assign weights to the different solutions. Matrix
element analyses ``naturally'' integrate over unconstrained variables.

For the most recent measurements presented here, both CDF and D\O\
consider only events containing two well identified electrons or
muons. Candidate events must have two oppositely charged leptons with
typically $\ET\ge$20(15)\,\gev\ in case of CDF (D\O) and at least two
jets with $\ET\ge$15(20)\,\gev.  The required amount of missing
transverse energy is typically higher than in the lepton-jets channel,
at least $\MET\ge$25(35)\,\gev.  Additional cuts are applied based on
the angle between the \MET\ vector and the transverse direction of the
leptons and jets, as well as on further topological variables. For
$ee$ and $\mu\mu$ events, the \MET\ requirement is modified (or the
event is rejected in case of D\O) if the di-lepton invariant mass lies
within a given window around the $Z$ boson mass, in order to reduce
more effectively background events with $Z\to l^+l^-$ decays. The
background is dominated by the Drell-Yan process, di-boson
contributions like $WW$+2 jets, and $W$+3 jet events where one jet was
misidentified as a lepton. The S/B ratios range from $\sim$2 for $\ge
0$ $b$-tags and $\sim$20 for $\ge1$ $b$-tag.

\begin{figure}[!t]
 \begin{center} 
  \includegraphics[width=.49\textwidth]{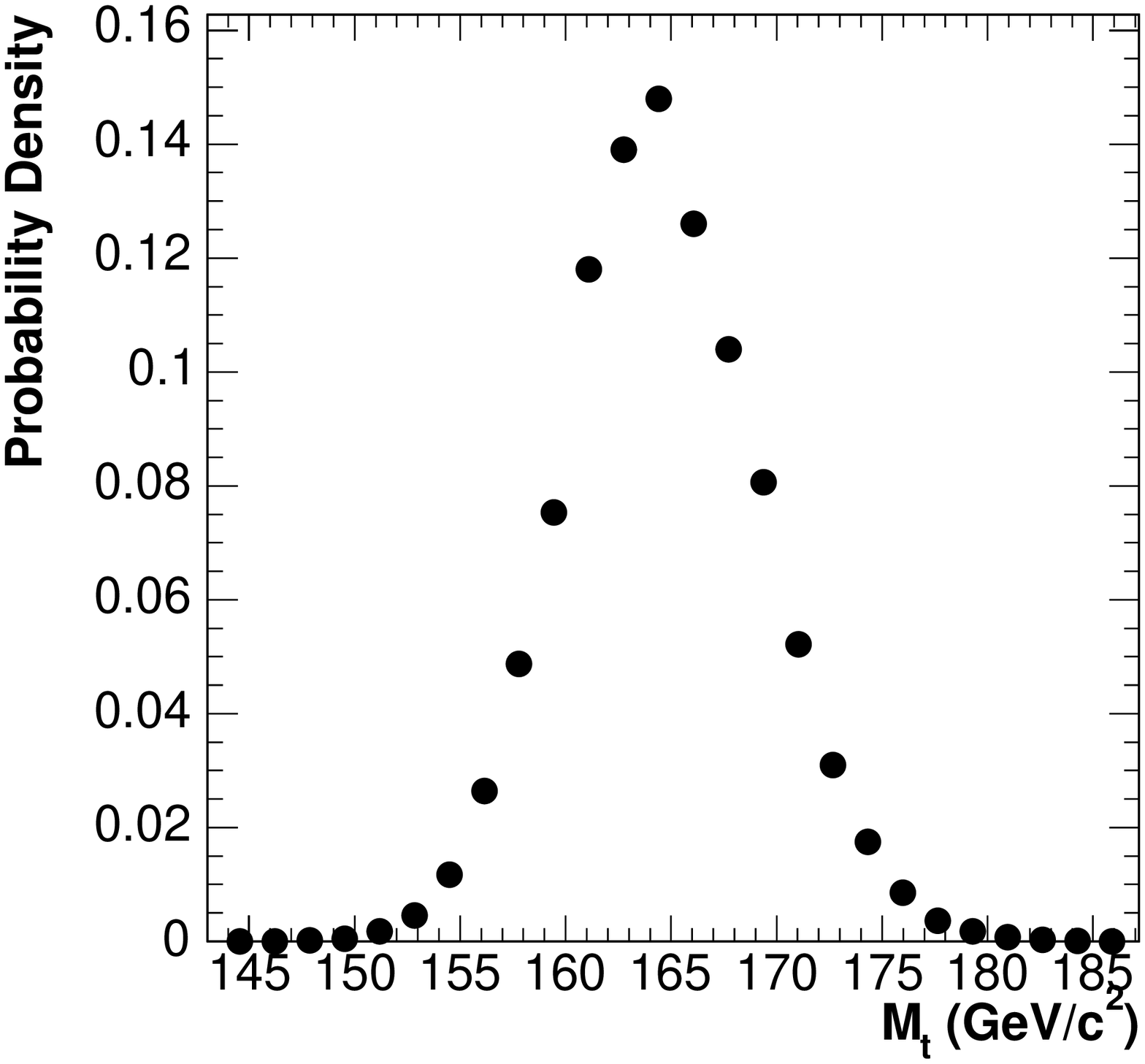}
  \includegraphics[width=.49\textwidth, height=54mm]{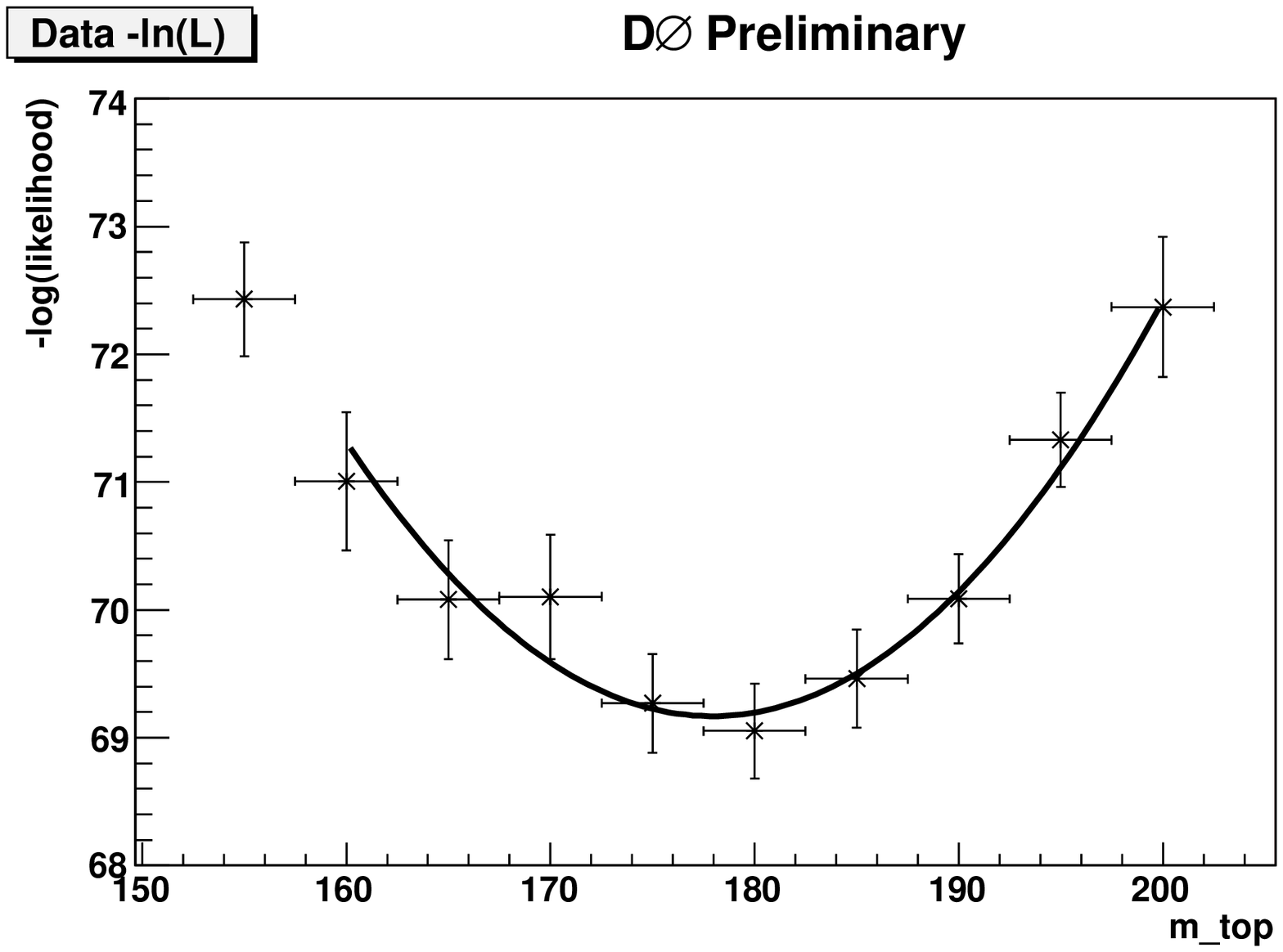}
 \vspace*{-1mm}
 \caption{\it Di-lepton channel results: Left: Joint likelihood
 vs. \Mt\ extracted by CDF with the Matrix Element Method
 (1030\,pb$^{-\mathit{1}}$). Right: Log likelihood curve obtained by
 D\O\ using the Matrix Weighting (835 pb$^{-\mathit{1}}$, $e\mu$
 events).
 \vspace*{-6mm}
\label{fig:mtop_dil} }
 \end{center}
\end{figure}

The best measurement in this channel is achieved by CDF using the
Matrix Element Method\cite{bib:CDF-DIL-ME}. The event probability
density is given by a linear combination of the probabilities for the
signal and the three major background processes using \Mt-dependent
weights from MC predictions. The individual probabilities are
calculated similarly to Eq.\,(\ref{eq:event-prob}), except for the
important difference that no JES {\em in situ} calibration is possible
on the basis of the signal process. Additional transfer functions for
the \ttbar\ transverse momentum are introduced using the
\ET\ of the sub-leading jets and the unclustered \ET,
in order to account for recoil effects induced by ISR. The
measurement is calibrated to account for limitations due to background
modeling and further simplifying assumptions. From a data set of
1030\,\invpb\ containing 78  candidate events ($27\pm5$
expected background), the likelihood curve shown in
Fig.\,\ref{fig:mtop_dil} (left) is extracted. The result obtained is
$\Mt=164.5\pm3.9^\mathrm{(stat.)}\pm3.5^\mathrm{(JES)}
\pm1.7^\mathrm{(syst.)}\,\gevcc=164.4\pm5.5\,\gevcc$ (3.3\% precision). 
By including background probabilities, the error is reduced by 15\%
compared to a measurement based on signal probability only. The
analysis was cross-checked using a subset of 30 events with at least
one $b$-tag, yielding
$\Mt=167.3\pm4.6^\mathrm{(stat.)}\pm3.3^\mathrm{(JES)}
\pm1.9^\mathrm{(syst.)}\,\gevcc=167.3\pm6.0\,\gevcc$ (3.6\% precision). 

The CDF and D\O\ collaborations have also applied the Template Method
in the di-lepton channel. D\O\ has employed two neutrino solution
weighting schemes\cite{bib:D0-DIL-TM}. The ``Neutrino Weighting
Method'' scans over the top quark mass \mt\ and the pseudorapidity of
the two neutrinos (ignoring the measured
\MET), and assigns weights based on the compatibility of the total
transverse neutrino energy with the observed \MET. For a given
\mt, the weights resulting from all neutrino $\eta$ assumptions and
two possible jet-quark assignments are summed. Detector resolutions
are taken into account by averaging the weights from repeated
calculations with input observables randomly smeared within their
resolutions. \Mt\ templates are formed using the sum of weights versus
\mt\ distributions. The analysis is performed with separate templates
for $ee$, $e\mu$ and $\mu\mu$ events, which are compared with the data
using a maximum likelihood fit. Using a sample of 1050\,\invpb\
containing 57 candidate events ($10.3^{+2.9}_{-3.7}$ expected
background), the result obtained is
$\Mt=172.5\pm5.8^\mathrm{(stat.)}\pm5.5^\mathrm{(syst.)}\,\gevcc
=172.5\pm8.0\,\gevcc$, which is the best single D\O\ measurement in
this channel (4.6\% precision).

Another scheme called ``Matrix Weighting Method'' assumes a value for
\mt\ and calculates the (at most) four corresponding neutrino solutions
given the $W$ mass, the jet and charge lepton momenta and \MET. A
weight is assigned based on the agreement of matrix element
predictions for the charged lepton \pT\ with the observed one.
Templates are built using the \mt\ values giving the maximum event
weight, and compared with the data using a binned maximum likelihood
fit. Fig.\,\ref{fig:mtop_dil} (right) shows the log likelihood curve
thus obtained.  Using 28 signal candidates in the $e\mu$-channel
($4.4^{+2.6}_{-1.4}$ expected background) selected in a sample of
835\,\invpb, D\O\ extracts $\Mt=177.7\pm8.8^\mathrm{(stat.)}\
^{+3.7}_{-4.5}\, ^\mathrm{(syst.)}\,\gevcc =177.7\pm9.7\,\gevcc$
(5.5\% precision).

CDF has also used the Neutrino Weighting Method in early Run-II as
well as a weighting scheme which scans the neutrino's azimuth
angle. Here we report on a more recent measurement known as ``Full
Kinematic Method''\cite{bib:CDF-DIL-TM}. The analysis assumes that the
distribution of the longitudinal momentum $p_z(\ttbar)$ of the
\ttbar\ system is a zero-centered Gaussian with $195\,\gevc$ width, as
indicated by MC simulations and supported by lepton-jets data. Studies
show that the $p_z(\ttbar)$ distribution has no mass dependence and is
equal for \ttbar\ and background.  Given $p_z(\ttbar)$ and using the
known $b$ and $W$ masses, one can solve the kinematic equations
numerically. The finite resolution is taken into account by smearing
the $b$-quark energies, \MET\ and $p_z(\ttbar)$ within the expected
uncertainties, and repeatedly solving the equations.  From the
resulting \mt\ distribution, the most probable value is taken to build
templates separately for events with and without $b$-tag. A maximum
likelihood fit to a 1.2\,\invfb\ data set with 70 candidates ($26\pm6$
background) yields $\Mt=169.1^{+5.2}_{-4.9}\
^\mathrm{(stat.)}\pm2.9^\mathrm{(JES)}\pm1.0^\mathrm{(syst.)}\,\gevcc
=169.1\pm5.9\,\gevcc$ (3.5\% precision). This is the most precise
template-based measurement in the di-lepton channel to date.

\section{Measurements in the All-Jets Channel} \label{sec:alljets}

Measurements in the all-jets channel are motivated by the high
branching ratio of 44\% and the complete reconstruction of the top
quarks, relying only on hadronic jets. The final state has
well-defined kinematics because no neutrino appears.  The channel is
challenging due to the huge background contamination and the large
combinatorial jet-quark ambiguity. Making no flavor requirement and
treating top-antitop permutations and the $W$ di-jet permutations
equally gives 90 combinations.

So far, only the CDF collaboration has reported measurements in this
channel.  The expected multi-jets final state has spherical topology
and well-balanced visible energy. The selection therefore requires
exactly six well-contained jets with $\ET>15\,\gev$ and a missing \ET\
significance $\MET/\sqrt{\sum\ET}<3\,\gev^\frac{1}{2}$. Events
containing high \pT\ electrons or muons are rejected.  Further cuts on
kinematic and topological variables are applied to purify the sample.
The remaining background is dominated by QCD multi-jet events ($bb4q$,
$6q$).

The first published Run-II measurement in the all-jets channel is
based on the Ideogram Method\cite{bib:CDF-AJ-Ideogramm}. The selection
used in this analysis yields S/B$\sim$1/25 without $b$-tagging
(compared to $\sim$1/3500 at trigger level) and S/B$\sim$1/5 including
$b$-tag information.  Similar to the D\O\ lepton-jets ideogram
analysis described in Sec.\ref{sec:lepjets}, individual signal
probability densities for the right and wrong jet-quark permutations
as well as background probability densities are formed, and weights
are assigned using the fit $\chisq$ and a $b$-tag probability measure.
To improve the S/B discrimination in the kinematic fit, the
probabilities are expanded in two dimensions using the invariant
masses of both the top and the antitop quark. For the signal they are
indistinguishable and expected to peak at the ``right'' value for \Mt,
but for background events at least one peaks at too low values. The
sample likelihood allows for a simultaneous optimization of both \Mt\
and the sample purity, because the QCD background cross sections are
not well known.  The result obtained using 290 $b$-tagged signal
candidates in a 310\,\invpb\ sample is
$\Mt=177.1\pm4.9^\mathrm{(stat.)}\pm4.3^\mathrm{(JES)}
\pm1.9^\mathrm{(syst.)}\,\gevcc=177.1\pm6.8\,\gevcc$
(3.8\% precision).

\begin{figure}[!t]
 \begin{center} 
 \includegraphics[width=.48\textwidth]{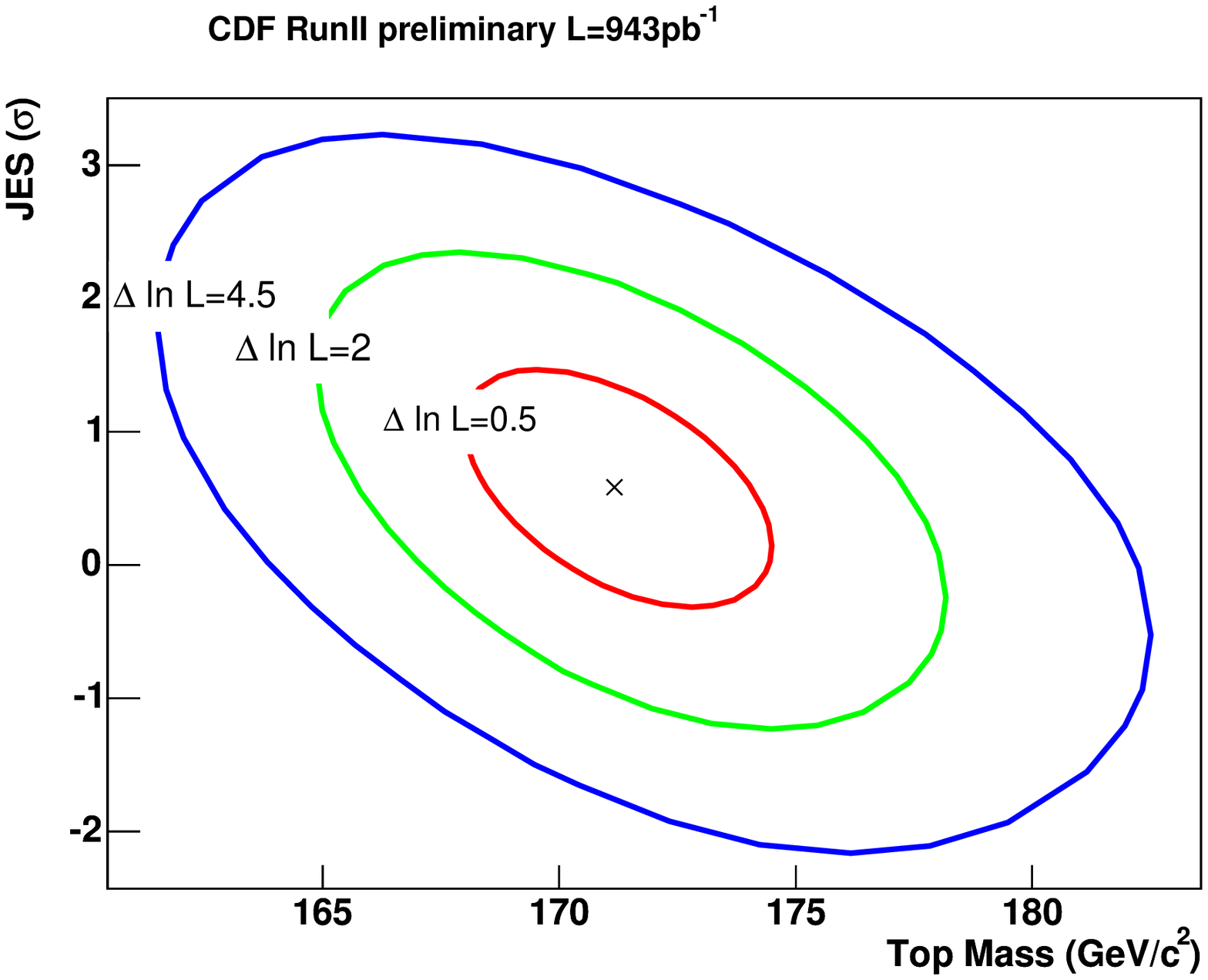} \hspace*{1mm}
 \includegraphics[width=.48\textwidth]{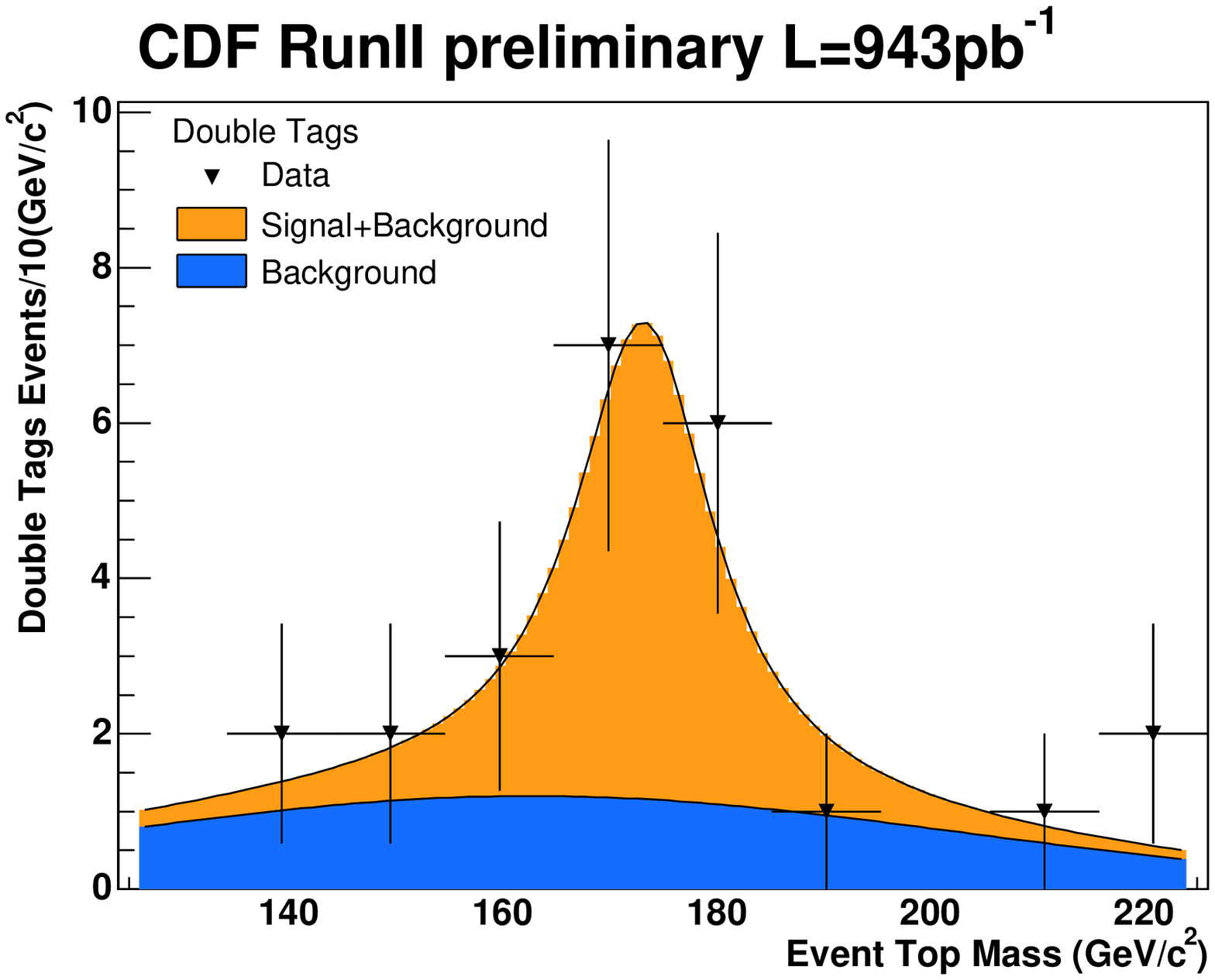} 
  \vspace*{-1mm}
  \caption{\it Left: Likelihood contours extracted by CDF from 943
  pb$^{-\mathit{1}}$ data using the Template Method with JES in situ
  calibration. Right: Top quark mass distribution in data and
  simulation for doubly tagged events.
  \vspace*{-6mm}
  \label{fig:mtop_alljets}}
 \end{center}
\end{figure}

The precision in the all-jets channel is greatly improved in a recent
template analysis\cite{bib:CDF-AJ-TM2} due to the adoption of the JES
{\em in situ} technique and pushing the S/B ratio to $\sim$1. The
latter was achieved using a novel neural network approach and by
considering samples with one and $\ge2$ $b$-tags separately.  The
signal templates are obtained using matrix element calculations and
transfer functions, the background probabilities are given by a data
driven model using the 0 $b$-tag sample, which has negligible signal
fraction.  Priors for the JES and the number of observed and
background events are used.  Fig.\,\ref{fig:mtop_alljets} (left) shows
the likelihood contours extracted from 943\,\invpb\ data containing 72
signal candidates ($\sim$44 estimated background). The good S/B ratio
achieved in this channel is illustrated in
Fig.\,\ref{fig:mtop_alljets} (right), which shows the doubly
$b$-tagged sample together with the fitted signal and background
templates. The analysis yields
$\Mt=171.1\pm2.8^\mathrm{(stat.)}\pm2.4^\mathrm{(JES)}
\pm2.1^\mathrm{(syst.)}\,\gevcc=171.1\pm4.3\,\gevcc$
(2.5\% precision). The JES uncertainty is much reduced compared to a
traditional one-dimensional template analysis\cite{bib:CDF-AJ-xsec-m}
based on 1020\,\invpb\ data containing 772 $b$-tagged candidate events,
which yields $\Mt=174.0\pm2.2^\mathrm{(stat.)}\pm4.5^\mathrm{(JES)}
\pm1.7^\mathrm{(syst.)}\,\gevcc=174.0\pm5.3\,\gevcc$
(3.0\% precision).

\section{Systematic Uncertainties} \label{sec:systematics}

So far, the systematic uncertainties of \Mt\ in all channels are
dominated by contributions from the JES. The di-lepton channel has the
biggest JES uncertainty because no {\em in situ} calibration is
performed here.  Other significant sources are primarily related to
the MC simulation. For the best results presented in this report,
these are the modeling of gluon ISR and FSR (particularly in the
all-jets and lepton-jets channel), the proton-parton density function,
the hadronization model, and the modeling of the background (specially
in the all-jets channel). The individual contributions are
$\sim$1\,\gevcc\ or less, and are expected be the limiting factor in
the precision of \Mt\ at the end of Run-II.

\section{Tevatron Combination} \label{sec:comb}

\begin{figure}[!t]
 \begin{center}
 \includegraphics[width=.49\textwidth]{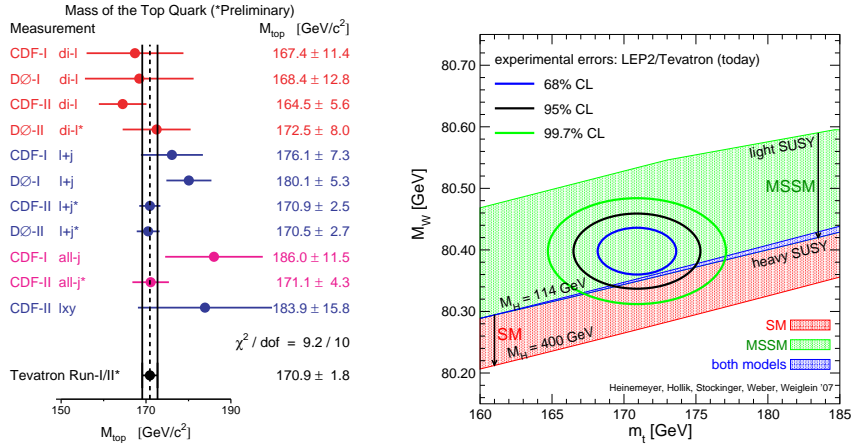}
 \includegraphics[width=.49\textwidth]{fig_mw_mt_mar07_1.eps}
 \vspace*{-2mm}
 \caption{\it Left: \Mt\ measurements used to calculate the current
 world average.  Right: \Mt\ versus \MW\ in the SM and the MSSM
 compared with the present measurements shown as confidence level
 contours\cite{bib:heinemeyer06}.
 \vspace*{-6mm}
 \label{fig:comb} }
 \end{center}
\end{figure}

CDF and D\O\ have updated the combination of their best results
achieved in each channel\cite{bib:tevcomb07} (see
Fig.\,\ref{fig:comb}, left), which includes the most precise
measurements reported in Sec.~\ref{sec:lepjets}, \ref{sec:dilep} and
\ref{sec:alljets}. Also the result from the 
Decay Length Technique is considered since its experimental
systematics are largely uncorrelated with those of other methods.
Taking all correlations between the systematic uncertainties properly
into account, the new world average value obtained is
$\Mt=170.9\pm1.1^\mathrm{(stat.)}\pm1.5^\mathrm{(syst.)}\,\gevcc
=170.9\pm1.8\,\gevcc$. The $\chisq$/d.o.f. of 9.2/10 (51\%
probability) indicates a good agreement among all measurements. New
\Mt\ averages are also calculated individually for each channel:
$172.2\pm4.1\,\gevcc$ (all-jets), $171.2\pm1.9\,\gevcc$ (lepton-jets),
$163.2\pm4.5\,\gevcc$ (di-lepton). The results are consistent given
their correlations.

\section{Conclusions} \label{sec:sum}

The CDF and D\O\ collaborations have established a robust top quark
mass measurement program based on a variety of techniques applied to
different \ttbar\ final states. The previous Run-I errors have been
reduced by a factor of 2-3 in each decay channel.  An important
achievement is the reduction of the JES uncertainty due to {\em in
situ} calibration, which is a reason why the all-jets channel has
become competitive. The new world average value for the top quark mass
is $\Mt=170.9\pm1.8\,\gevcc$, which corresponds to a precision of
1.1\%. The effect of \Mt\ and the recently updated \MW\
measurement\cite{bib:proc-hays} on the mass of the Higgs boson is
shown in Fig.\,\ref{fig:comb} (right). The uncertainties translate to
a $\sim$30\% constraint for \MH. With full Run-II data, the
uncertainty in \Mt\ may be even pushed to 1\,\gevcc, which is also
expected after 5-10 years of LHC operation. The top quark mass might
thus be the lasting legacy of the Tevatron.

\section*{Acknowledgments}
I would like to thank the conference organizers for the kind
invitation, and also my colleagues from the CDF and D\O\
collaborations for their help in preparing the talk.  I am also
grateful to the Max Planck Society and the Alexander von Humboldt
Foundation for their support.


%
\end{document}